\documentclass[doublecol]{epl2}
\usepackage{graphicx}
\usepackage{amsmath}
\usepackage{amssymb}
\usepackage{bm}
\usepackage{epsfig}
\usepackage{bm}
\usepackage{indentfirst}

\title{Magnetoresistance of the heavy-fermion metal \chem{CeCoIn_5}}
\shorttitle{Magnetoresistance  of the heavy-fermion metal
\chem{CeCoIn_5}}

\author{V.R. Shaginyan\inst{1}\thanks {Email:
\email{vrshag@thd.pnpi.spb.ru}} \and K.G. Popov\inst{2}}
\shortauthor{V.R. Shaginyan and K.G. Popov}

\institute{\inst{1} Petersburg Nuclear
Physics Institute, Gatchina, 188300, Russia\\
\inst{2} Komi Science Center, Ural Division, RAS, Syktyvkar, 167982,
Russia}
 {Quantum
phase transitions}

\abstract{The magnetoresistance (MR) of \chem{CeCoIn_5} is notably
different from that expected for orbital MR due to the Lorentz force
and described by Kohler's rule which holds in many conventional
metals. We show that a pronounced crossover from negative to
positive MR of \chem{CeCoIn_5} that occurs at elevated temperatures
is determined by the dependence of the effective mass $M^*(B,T)$ on
both magnetic field $B$ and temperature $T$. Thus, the crossover is
regulated by the universal behavior of $M^*(B,T)$ observed in
heavy-fermion metals. This behavior is exhibited by $M^*(B,T)$ when
a strongly correlated electron system transits from the Landau Fermi
liquid behavior induced by the application of magnetic field to the
non-Fermi liquid behavior taking place at rising temperatures. Our
calculations of MR are in good agreement with facts and reveal new
scaling behavior of MR.}

\begin{document}

\maketitle

An explanation of the rich and striking behavior of strongly
correlated electron system in heavy fermion (HF) metals is, as years
before, among the main problems  of the condensed matter physics.
One of the most interesting and puzzling issues in the research of
HF metals is the anomalous normal-state transport properties. HF
metals show a number of distinctive transport properties, among
which is the magnetoresistance (MR) of HF metals. Measurements of MR
on \chem{CeCoIn_5} \cite{pag,mal} have shown that this is notably
different from that expected for weak-field orbital MR and described
by Kohler's rule which holds in many conventional metals, see e.g.
\cite{zim}. MR of \chem{CeCoIn_5} exhibits a crossover from negative
to positive MR that occurs in fixed magnetic fields $B$ with
increasing temperature $T$, so that at the high fields  and
relatively low temperatures MR becomes negative \cite{pag,mal}.

This crossover is hard to explain within the conventional Fermi
liquid theory for metals and in terms of Kondo systems
\cite{daybell} and therefore it is assumed that the crossover can be
attributed to some energy scales causing a change in character of
spin fluctuations with increasing the applied magnetic field
strength \cite{pag}. It is widely believed that such quantum
fluctuations becoming sufficiently strong suppress quasiparticles at
a quantum phase transition and when the system in question transits
from its Landau-Fermi liquid (LFL) regime to non-Fermi liquid (NFL)
behavior \cite{voj,loh,si}.

On the other hand, even early measurements carried out on HF metals
gave evidences in favor of the existence of quasiparticles. For
example, the application of magnetic field $B$ restores LFL regime
of HF metals which in the absence of the field demonstrates NFL
behavior. In that case the empirical Kadowaki-Woods ratio $K$ is
conserved, $K=A(B)/\gamma_0^2(B)\propto A(B)/\chi^2(B)=const$
\cite{kadw,tky} where $C/T=\gamma_0$, $C$ is the heat capacity,
$\chi$ is magnetic susceptibility and $A(B)$ is the coefficient
determining the temperature dependence of the resistivity
$\rho=\rho_0+A(B)T^2$. Here $\rho_0$ is the residual resistance. The
observed conservation of $K$ can be hardly accounted for within
scenarios when quasiparticles are suppressed, for there is no reason
to expect that $\gamma_0(B)$, $\chi(T)$, $A(B)$ and other transport
and thermodynamic quantities like the thermal expansion coefficient
$\alpha(B)$ are affected by the fluctuations or localization in a
correlated fashion.

Quasiparticles were observed in LFL regime in measurements of
transport properties on \chem{CeCoIn_5} \cite{pag1}. While it is
extensively accepted that the NFL behavior is determined by the
critical fluctuations, Kondo lattice \cite{voj,loh,si} and multiple
energy scales \cite{steg}, therefore in that scenario the crossover
region has to be formed by the fluctuations and scales rather then
by quasiparticles. Analyzing thermodynamic quantities, it was shown
that quasiparticles exist in both the LFL and the crossover regimes
when strongly correlated Fermi systems such as HF metals
\cite{ckz,khod,obz,shag1,shag2} or two-dimensional \chem{^3He}
\cite{shag3} transit from its LFL to NFL behavior. Therefore it is
of crucial importance to verify whether quasiparticles characterized
by their effective mass $M^*$ still exist and determine the
transport properties of HF metals in the crossover region. As we
will see, measurements of MR in the crossover region can present
indicative data on the availability of quasiparticles. Fortunately,
such measurements of MR were carried out on \chem{CeCoIn_5} when the
system transit from the LFL  to NFL behavior at elevated
temperatures and fixed magnetic fields \cite{pag,mal}.

In this Letter we analyze MR of \chem{CeCoIn_5} and show that the
crossover from negative to positive MR that occurs at elevated
temperatures and fixed magnetic fields can be well captured
utilizing fermion condensation quantum phase transition (FCQPT)
based on the quasiparticles paradigm \cite{khs,obz,ams,volovik}. We
demonstrate that crossover is regulated by the universal behavior of
the effective mass $M^*(B,T)$ observed in many heavy-fermion metals
and is exhibited by $M^*(B,T)$ when HF metal transits from the LFL
behavior induced by the application of magnetic field to NFL
behavior taking place at rising temperatures. Our calculations of MR
are in good agreement with facts and allow us to reveal new scaling
behavior of MR. Thus, we show that the transport properties are
mainly determined by quasiparticles rather then by the critical
fluctuations, Kondo lattice and energy scales which are expected to
arrange the behavior in the transition region.

To study universal low temperature features of HF metals, we use the
model of homogeneous heavy-fermion liquid with the effective mass
$M^*(T,B,x)$, where the number density $x=p_F^3/3\pi^2$ and $p_F$ is
the Fermi momentum \cite{land}. This permits to avoid complications
associated with the crystalline anisotropy of solids \cite{shag1}.
We first outline the case when at $T\to 0$ the heavy-electron liquid
behaves as LFL and is brought to the LFL side of FCQPT by tuning a
control parameter like $x$. At elevated temperatures the system
transits to the NFL state. The dependence $M^*(T,x)$ is governed by
Landau equation \cite{land}
\begin{equation}
\frac{1}{M^*(T,x)}=\frac{1}{M}+\int\frac{{\bf p}_F{\bf p}}{p_F^3}F
({\bf p_F},{\bf p})\frac{\partial n({\bf
p},T,x)}{\partial{p}}\frac{d{\bf p}}{(2\pi)^3},\label{LQ}
\end{equation}
where $n({\bf p},T,x)$ is the distribution function of
quasiparticles and $F({\bf p}_F,{\bf p})$ Landau interaction
amplitude. At $T=0$, eq. \eqref{LQ} reads \cite{land}
$M^*/M=1/(1-N_0F^1(p_F,p_F)/3)$. Here $N_0$ is the density of states
of a free electron gas,  $F^1(p_F,p_F)$ is the $p$-wave component of
Landau interaction amplitude $F$. Taking into account that $x
=p_F^3/3\pi^2$, we rewrite the amplitude as $F^1(p_F,p_F)=F^1(x)$.
When at some critical point $x=x_{FC}$, $F^1(x)$ achieves certain
threshold value, the denominator tends to zero and the system
undergoes FCQPT related to divergency of the effective mass
\cite{khs,volovik,obz}
\begin{equation}\label{zui2}
\frac{M^*(x)}{M}=A+\frac{B}{x_{FC}-x},
\end{equation}
where $M$ is the bare mass, eq. (\ref{zui2}) is valid in both 3D and
2D cases, while the values of factors $A$ and $B$ depend on the
dimensionality. The approximate solution of eq. \eqref{LQ} is of the
form \cite{shag2}
\begin{eqnarray}
\frac{M}{{M^* (T)}}&=&\frac{M}{{M^*(x)}}+\beta f(0)\ln\left\{
{1+\exp(-1/\beta)}\right\}\nonumber \\
&+&\lambda _1\beta^2+\lambda_2 \beta^4 + ...,\label{zui1}
\end{eqnarray}
where $\lambda_1>0$ and $\lambda_2<0$ are constants of order unity,
$\beta=TM^*(T)/p_F^2$ and $f(0)\sim F^1(x_{FC})$. It follows from
eq. (\ref{zui1}) that the effective mass $M^*$ as a function of $T$
and $x$ reveals three different regimes at growing temperature. At
the lowest temperatures we have LFL regime with $M^*(T,x)\simeq
M^*(x)+aT^2$ with $a<0$ since $\lambda_1>0$. The effective mass as a
function of $T$ decays down to a minimum and after grows, reaching
its maximum $M^*_M$ at some temperature $T_M(x)$ then subsequently
diminishing as $T^{-2/3}$ \cite{ckz,obz}. Moreover, the closer is
the number density $x$ to its threshold value $x_{FC}$, the higher
is the rate of the growth. The peak value $M^*_M$ grows also, but
the maximum temperature $T_M$ lowers. Near this temperature the last
"traces" of LFL regime disappear, manifesting themselves in the
divergence of above low-temperature series and substantial growth of
$M^*(x)$. Numerical calculations based on eqs. (\ref{LQ}) and
(\ref{zui1}) show that at rising temperatures $T>T_{1/2}$ the linear
term $\propto \beta$ gives the main contribution and leads to new
regime when eq. (\ref{zui1}) reads $M/M^*(T)\propto\beta $ yielding
\begin{equation}\label{r2}
    M^*(T) \propto T^{-1/2}.
\end{equation} We remark that eq. \eqref{r2} ensures that at $T\geq
T_{1/2}$ the resistivity behaves as $\rho(T)\propto T$ \cite{obz}.
\begin{figure}[!ht]
\begin{center}
\includegraphics [width=0.37\textwidth]{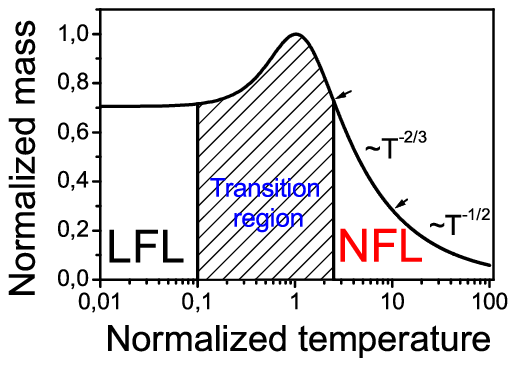}
\vspace*{-0.3cm}
\end{center}
\caption{Schematic plot of the normalized effective mass
$M^*_N=M^*(T/T_M)/M^*_M$ ($M^*_M$ is its maximal value at $T=T_M$)
versus the normalized temperature $T_N=T/T_M$. Several regions are
shown. First goes the LFL regime ($M^*_N(T_N)\sim$ const) at $T_N\ll
1$, then the transition regime (the hatched area) where $M^*_N$
reaches its maximum. At elevated temperatures $T^{-2/3}$ regime
occurs followed by $T^{-1/2}$ behavior, see eq.
(\ref{r2}).}\label{MT}
\end{figure}

Near the critical point $x_{FC}$, with $M/M^*(x\to x_{FC})\to0$, the
behavior of the effective mass changes dramatically because the
first term on the right-hand side of eq. \eqref{zui1} vanishes, the
second term becomes dominant, and the effective mass is determined
by the homogeneous version of \eqref{zui1} as a function of $T$. As
a result, the scale $M/M^*$ vanishes and we get to scale $M^*$ in
$M^*_M$ and $T$ in $T_M$. These scales can be viewed as natural
ones. The schematic plot of the normalized effective mass
$M^*_N=M^*/M^*_M$ versus normalized temperature $T_N=T/T_M$ is
reported fig. \ref{MT}. In fig. \ref{MT} both $T^{-2/3}$ and
$T^{-1/2}$ regimes are marked as NFL ones since the effective mass
depends strongly on temperature. The temperature region $T\simeq
T_M(x)$ signifies the crossover between the LFL regime with almost
constant effective mass and NFL behavior, given by $T^{-2/3}$
dependence. Thus temperatures $T\sim T_M$ can be regarded as the
crossover between LFL and NFL regimes. It turns out that $M^*(T,x)$
in the entire $T\leq T^{-1/2}$ range can be well approximated by a
simple universal interpolating function \cite{obz,shag2,ckz}. The
interpolation occurs between the LFL ($M^*\propto T^2$) and NFL
($M^*\propto T^{-1/2}$, see eq. \eqref{r2}) regimes thus describing
the above crossover. Introducing the dimensionless variable
$y=T_N=T/T_M$, we obtain the desired expression
\begin{equation}
\frac{M^*(T/T_N,x)}{M^*_M} = {M^*_N(y)}\approx
\frac{M^*(x)}{M^*_M}\frac{1+c_1y^2}{1+c_2y^{8/3}}. \label{UN2}
\end{equation}
Here $M^*_N(y)$ is the normalized effective mass,  $c_1$ and $c_2$
are parameters, obtained from the condition of best fit to
experiment. To correct the behavior of $M^*_N(y)$ at rising
temperatures we add a term to eq. \eqref{UN2} and obtain
\begin{equation}
M^*_N(y)\approx\frac{M^*(x)}{M^*_M}\left[\frac{1+c_1y^2}{1+c_2y^{8/3}}
+c_3\frac{\exp(-1/y)}{\sqrt{y}}\right], \label{HC28}\end{equation}
where $c_3$ is a parameter. The last term on the right hand side of
eq. \eqref{HC28} makes $M^*_N$ satisfy eq. \eqref{r2} at rising
temperatures $T/T_M>2$.

At small magnetic fields $B$ (that means that the Zeeman splitting
is small), the effective mass does not depend on spin variable and
$B$ enters eq. \eqref{LQ} as $B\mu_B/T$ making $T_M\propto B\mu_B$
where $\mu_B$ is the Bohr magneton \cite{obz,shag2}. The application
of magnetic field restores the LFL behavior, and at $T=0$ the
effective mass depends on $B$ as \cite{ckz,obz}
\begin{equation}\label{B32}
M^*(B)\propto (B-B_{c0})^{-2/3},
\end{equation} where $B_{c0}$ is the
critical magnetic field driving both HF metal to its magnetic field
tuned QCP and corresponding N\'eel temperature toward $T=0$. In some
cases $B_{c0}=0$. For example, the HF metal \chem{CeRu_2Si_2} is
characterized by $B_{c0}=0$ and shows neither evidence of the
magnetic ordering or superconductivity nor the LFL behavior down to
the lowest temperatures \cite{takah}. In our simple model $B_{c0}$
is taken as a parameter. We conclude that under the application of
magnetic filed the variable
\begin{equation}\label{YTB}
y=T/T_M\propto \frac{T}{\mu_B(B-B_{c0})}
\end{equation}
remains the same and the normalized effective mass is again governed
by eqs. \eqref{UN2} and \eqref{HC28} which are the final result of
our analytical calculations. We note that the obtained results are
in agreement with numerical calculations \cite{obz,ckz}.

The normalized effective mass $M^*_N(y)$ can be extracted from
experiments on HF metals. For example, $M^*(T,B)\propto
C(T)/T\propto S(T)/T\propto \chi_{AC}(T)$, where $S(T)$ is the
entropy, $C(T)$ is the specific heat and $\chi_{AC}(T)$ is ac
magnetic susceptibility. If the corresponding measurements are
carried out at fixed magnetic field $B$ (or at fixed both the
concentration $x$ and $B$) then, as it seen from fig. \ref{MT}, the
effective mass reaches the maximum at some temperature $T_M$. Upon
normalizing both the effective mass by its peak value $M^*_M$ at
each field $B$  and the temperature by $T_M$, we observe that all
the curves have to merge into single one, given by eqs. (\ref{UN2})
and \eqref{HC28} thus demonstrating a scaling behavior.

\begin{figure} [! ht]
\begin{center}
\includegraphics [width=0.47\textwidth]{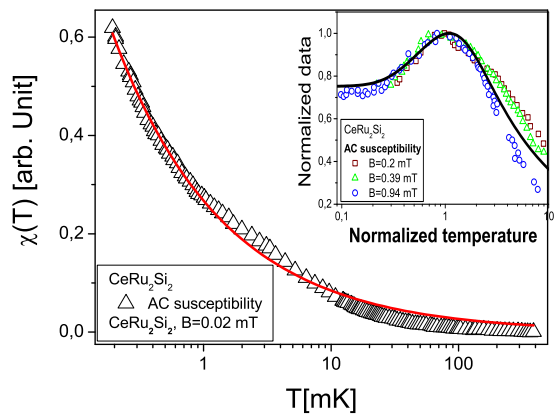}
\vspace*{-1.0cm}
\end{center}
\caption{Temperature dependence of the $ac$ susceptibility
$\chi_{AC}$  for \chem{CeRu_2Si_2}. The solid curve is a fit for the
data shown by the triangles at $B=0.02$ mT and represented by the
function $\chi(T)=a/\sqrt{T}$ given by eq. \eqref{r2} with $a$ being
a fitting parameter. Inset shows the normalized effective mass
versus normalized temperature $y$ extracted from $\chi_{AC}$
measured at different fields as indicated in the inset \cite{takah}.
The solid curve traces the universal behavior of $M^*_N(y)$
determined by eq. (\ref{UN2}). Parameters $c_1$ and $c_2$ are
adjusted to fit the average behavior of the normalized effective
mass $M^*_N(y)$.}\label{MRM}
\end{figure}

To verify eq. \eqref{r2}, we use measurements of $\chi_{AC}(T)$ in
\chem{CeRu_2Si_2} at magnetic field $B=0.02$ mT at which this HF
metal demonstrates the NFL behavior \cite{takah}. It is seen from
fig. \ref{MRM} that eq. \eqref{r2} gives good description of the
facts in the extremely wide range of temperatures. The inset of fig.
\ref{MRM} exhibits a fit for $M^*_N(y)$ extracted from measurements
of $\chi_{AC}(T)$ at different magnetic fields, clearly indicating
that the function given by eq. \eqref{UN2} represents a good
approximation for $M^*_N(y)$.

\begin{figure} [! ht]
\begin{center}
\includegraphics [width=0.49\textwidth]{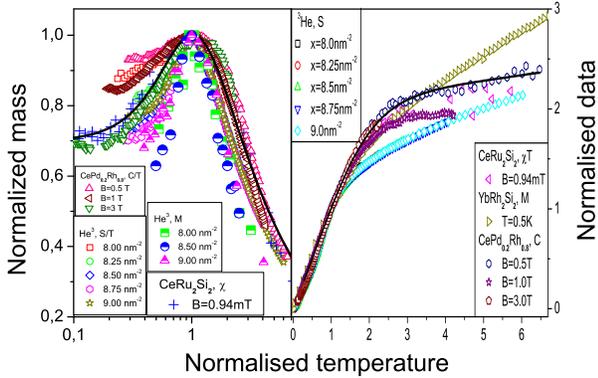}
\end{center}
\vspace*{-0.8cm} \caption{The left panel. The normalized effective
mass $M^*_N$ versus the normalized temperature $y=T/T_M$. The
behavior of $M^*_N$ is extracted from measurements of $S(T)/T$ and
magnetization $M$ on 2D \chem{^3He} \cite{he3}), from $ac$
susceptibility $\chi_{AC}(T)$ collected on \chem{CeRu_2Si_2}
\cite{takah} and from $C(T)/T$ collected on \chem{CePd_{1-x}Rh_x}
\cite{pikul}. The data are collected at different densities and
magnetic fields as shown in the left bottom corner. The solid curve
traces the universal behavior of the normalized effective mass
determined by eq. (\ref{UN2}). Parameters $c_1$ and $c_2$ are
adjusted for $\chi_{N}(T_N,B)$ at $B=0.94$ mT. The right panel. The
normalized specific heat $C(y)$ of \chem{CePd_{1-x}Rh_x} at
different magnetic fields $B$, normalized entropy $S(y)$ of
\chem{^3He} at different number densities $x$, and the normalized
$y\chi(y)$ at $B=0.94$ mT versus normalized temperature $y$ are
shown. The upright triangles depict the normalized `average'
magnetization $M+B\chi$ collected on \chem{YbRu_2Si_2} \cite{steg}.
The kink in all the data is clearly seen in the transition region
$y\geq 1$. The solid curve represents $yM^*_N(y)$ with parameters
$c_1$ and $c_2$ adjusted for magnetic susceptibility of
\chem{CeRu_2Si_2} at $B=0.94$ mT.}\label{f2}
\end{figure}

$M^*_N(y)$ extracted from the entropy $S(T)/T$ and magnetization $M$
measurements on the $^3$He film \cite{he3} at different densities
$x<x_{FC}$ is reported in the left panel of fig. \ref{f2}. In the
same panel, the data extracted from the heat capacity of the
ferromagnet \chem{CePd_{0.2}Rh_{0.8}} \cite{pikul} and the AC
magnetic susceptibility of the paramagnet \chem{CeRu_2Si_2}
\cite{takah} are plotted for different magnetic fields. It is seen
that the universal behavior of the normalized effective mass given
by eq. (\ref{UN2}) and shown by the solid curve is in accord with
the experimental facts. All 2D \chem{^3He} substances are located at
$x<x_{FC}$, where the system progressively disrupts its LFL behavior
at elevated temperatures. In that case the control parameter,
driving the system towards its critical point $x_{FC}$ is merely the
number density $x$. It is seen that the behavior of $M^*_N(y)$,
extracted from $S(T)/T$ and magnetization $M$ of 2D \chem{^3He}
looks very much like that of 3D HF compounds. In the right panel of
fig. \ref{f2}, the normalized data on $C(y)$, $S(y)$, $y\chi(y)$ and
$M(y)+y\chi(y)$ extracted from data collected on
\chem{CePd_{1-x}Rh_x} \cite{pikul} , \chem{^3He} \cite{he3},
\chem{CeRu_2Si_2} \cite{takah}, and \chem{YbRu_2Si_2} \cite{steg}
respectively are presented. Note that in the case of
\chem{YbRu_2Si_2}, the variable $y=B\mu_B/T_M$. As seen from eq.
\eqref{UN2}, this representation of the variable $y$ is correct, and
$B\mu_B$ makes sense of the variable, while the temperature is a
fixed parameter. All the data show a kink at $y\geq 1$ taking place
as soon as the system enters the transition region from the LFL
state to the NFL one. Again, we conclude that the presence of the
kink is mainly determined by the behavior of the effective mass at
the transition region rather then by the critical fluctuations or
Kondo scales \cite{steg}.

By definition, MR is given by \begin{equation}
\rho_{mr}(B,T)=\frac{\rho(B,T)-\rho(0,T)}{\rho(0,T)},\label{HC23}
\end{equation}
We apply eq. (\ref{HC23}) to study MR of strongly correlated
electron liquid versus temperature $T$ as a function of magnetic
field $B$. The resistivity $\rho(B,T)$ is
\begin{equation}
\rho(B,T)=\rho_0+\Delta\rho(B,T)+\Delta\rho_{L}(B,T),
\label{RBT}\end{equation} where $\Delta\rho(B,T)=c(M^*(B,T))^2T^2$,
and $c$ is a constant, and the classical contribution
$\Delta\rho_{L}(T,B)$ to MR due to orbital motion of carriers
induced by the Lorentz force obeys the Kohler's rule \cite{zim}:
\begin{equation}
\frac {\Delta\rho_{L}(B,T)}{\rho(0, T)}\simeq
F\left(\frac{\mu_BB}{\rho(0, T)}\right).\label{HC24}
\end{equation}
Function $F$ is determined by the details of metal. We note that
$\Delta\rho_{L}(B)$ $\ll\rho(0, T)$ as it is assumed in the
weak-field approximation. Suppose that the temperature is not very
low, so that $\rho_0\leq \Delta\rho(B=0,T)$, and $B\geq B_{c0}$.
Substituting (\ref{RBT}) in (\ref{HC23}), we find that
\begin{eqnarray}
\nonumber
&&\rho_{mr}(B,T)\simeq \frac{\Delta\rho_{L}(B,T)}{\rho(0, T)}\\
&+&cT^2 \frac{(M^*(B,T))^2-(M^*(0, T))^2}{\rho(0, T)}\label{HC25}.
\end{eqnarray}

Consider the qualitative behavior of MR described by eq.
(\ref{HC25}) as a function of $B$ at a certain temperature $T=T_0$.
In weak magnetic fields, when $T_0\geq T_{1/2}$ and the system
exhibits NFL regime (see fig. \ref{MT}), the main contribution to MR
is made by the term $\Delta\rho_{L}(B)$, because the effective mass
is independent of the applied magnetic field. Hence, $|M^*(B,
T)-M^*(0,T)|/M^*(0, T)\ll1$ and the leading contribution is made by
$\Delta\rho_{L}(B)$. As a result, MR is an increasing function of
$B$. When $B$ becomes so high that $T_M(B)\sim \mu_BB\sim T_0$, the
difference $(M^*(B, T)-M^*(0, T))$ becomes negative and MR as a
function of $B$ reaches its maximum value at $T_M(B)\sim T_0$. As
$B$ increases still further, when $T_M(B)>T_0$, the effective mass
$M^*(B,T)$ becomes a decreasing function of the magnetic field, as
follows from eq. (\ref{B32}). As $B$ increases,
\begin{equation}\frac{(M^*(B,T)-M^*(0,T))}
{M ^*(0, T)}\to -1,\label{HC25a}\end{equation} and the
magnetoresistance, being a decreasing function of $B$, is negative.

Now study the behavior of MR as a function of $T$ at a certain value
$B_0$ of magnetic field. At low temperatures $T\ll T_M(B_0)$, it
follows from eqs. (\ref{UN2}) and (\ref{B32}) that
\begin{equation}\frac{M^*(B_0)}{M^*(T)}
\ll1,\label{HC25b}\end{equation} and it follows from eq.
\eqref{HC25a} that $\rho_{mr}(B_0,T)\sim-1$, because
$\Delta\rho_{L}(B)/\rho(0,T)\ll1$. We note that $B_0$ must be
relatively high to guarantee that $M^*(B_0)/M^*(T)$ $\ll1$. As the
temperature increases, MR increases, remaining negative. At $T\simeq
T_M(B_0)$, MR is approximately zero, because $M^*(B_0)\simeq M^*(T)$
and $\rho(B_0,T)\simeq\rho(0,T)$ at this point. This allows us to
conclude that the change of the temperature dependence of
resistivity $\rho(B_0,T)$ from quadratic to linear manifests itself
in the transition from negative MR to positive. One can also say
that the transition takes place when the system transits from the
LFL behavior to the NFL one. At $T\geq T_M(B_0)$, the leading
contribution to MR is made by $\Delta\rho_{L}(B_0)$ and MR reaches
its maximum. At $T_M(B_0)\ll T$, MR is a decreasing function of the
temperature, because
\begin{equation}\frac{|M^*(B,T)-M^*(0, T)|}{M^*(0,T)}
\ll1,\label{HC25b}\end{equation} and $\rho_{mr}(B_0,T)\ll1$.

The both transitions \cite{shag_mr} (from positive MR to negative MR
with increasing $B$ at a fixed temperature $T$ and from negative MR
to positive MR with increasing $T$ at a fixed value of $B$) have
been detected in measurements of the resistivity of \chem{CeCoIn_5}
in a magnetic field \cite{pag}.

\begin{figure} [! ht]
\begin{center}
\includegraphics [width=0.47\textwidth]{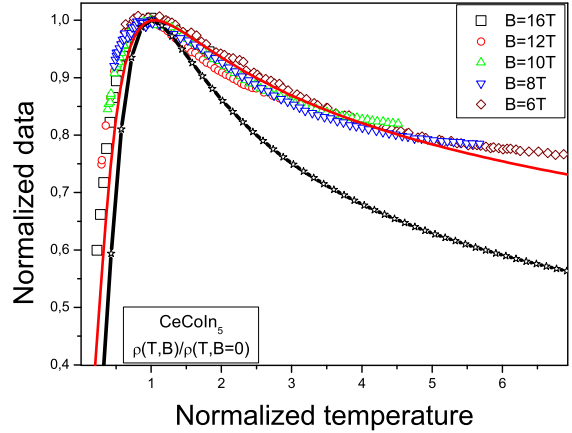}
\end {center}\vspace{-0.8cm}
\caption {The normalized ratio $R^{\rho}_N(y)$ given by eq.
\eqref{HC27} versus normalized temperature $y=T/T_{\rm Rm}$. The
normalized ratio were extracted from MR shown in fig. \ref{MRT} and
collected on \chem{CeCoIn_5} at fixed magnetic fields $B$ \cite{pag}
listed in the right upper corner. The starred  line represents our
calculations based on eqs. \eqref{UN2} and \eqref{HC27} with the
parameters extracted from $ac$ susceptibility of $\rm CeRu_2Si_2$
(see the caption to fig. \ref{MRM}). The solid line displays our
calculations based on eqs. \eqref{HC28} and \eqref{HC27}; only one
parameter was used to fit the data, while the other were extracted
from the $ac$ susceptibility measured on
\chem{CeRu_2Si_2}.}\label{MRTU}
\end{figure}

Let us turn to quantitative analysis of MR. As it was mentioned
above, we can safely assume that the classical contribution
$\Delta\rho_{L}(B,T)$ to MR is small as compared with
$\Delta\rho(B,T)$. This fact allows us to make our analysis and
results transparent and simple since the behavior of
$\Delta\rho_{L}(B_0)$ is not known in the case of HF metals.
Consider the ratio $R^{\rho}=\rho(B,T)/\rho(0,T)$ and assume for a
while that the residual resistance $\rho_0$ is small in comparison
with the temperature dependent terms. Taking into account eq.
\eqref{RBT} and that $\rho(0,T)\propto T$, we obtain from eq.
\eqref{HC25} that
\begin{equation}R^{\rho}=\rho_{mr}+1=
\frac{\rho(B,T)}{\rho(0, T)}\propto T(M^*(B,T))^2.\label{HC26}
\end{equation} It follows from eqs.  \eqref{UN2} and \eqref{HC26}
that the ratio $R^{\rho}$ reaches its maximum value $R^{\rho}_M$ at
some temperature $T_{\rm Rm}\sim T_M$. If the ratio is measured in
terms of its maximum value $R^{\rho}_M$ and $T$ is measured in terms
of $T_{\rm Rm}\sim T_M$ then it is seen from eqs. \eqref{UN2},
\eqref{HC28} and \eqref{HC26} that the normalized ratio
\begin{equation}R^{\rho}_N(y)=
\frac{R^{\rho}(B,T)}{R^{\rho}_M(B)}\simeq y(M^*_N(y))^2\label{HC27}
\end{equation} becomes a universal function of the only variable
$y=T/T_{\rm Rm}$. To verify eq. \eqref{HC27}, we use MR obtained in
measurements on CeCoIn$_5$, see fig. 1(b) of Ref. \cite{pag}. The
results of the normalization procedure of MR are reported in fig.
\ref{MRTU}. It is clearly seen that the data collapse into the same
curve, indicating that  MR well obeys the scaling behavior given by
eq. \eqref{HC27}. This scaling behavior obtained directly from the
experimental facts is a vivid proof that MR is predominantly
governed by the effective mass $M^*(B,T)$.

Now we are in position to calculate $R^{\rho}_N(y)$ given by eq.
\eqref{HC27}. Using eq. \eqref{UN2} to parametrize $M^*_N(y)$, we
extract parameters $c_1$ and $c_2$ from measurements of the magnetic
$ac$ susceptibility $\chi$ on $\rm CeRu_2Si_2$ \cite{takah} and
apply eq. \eqref{HC27} to calculate the normalized ratio. It is seen
that the calculations shown by the starred line in fig. \ref{MRTU}
start to deviate from the experimental facts at elevated
temperatures. To improve the description, we employ eq. \eqref{HC28}
which describes the behavior of the effective mass at elevated
temperatures in accord with eq. \eqref{r2} and ensures that at these
temperatures the resistance behaves as $\rho(T)\propto T$. In fig.
\ref{MRTU}, the fit of $R^{\rho}_N(y)$ by eq. \eqref{HC28} is shown
by the solid line. Constant $c_3$ is taken as a fitting parameter,
while the other were extracted from $ac$ susceptibility of
\chem{CeRu_2Si_2} as described in the caption to fig. \ref{MRM}.

\begin{figure} [! ht]
\begin{center}
\includegraphics [width=0.40\textwidth]{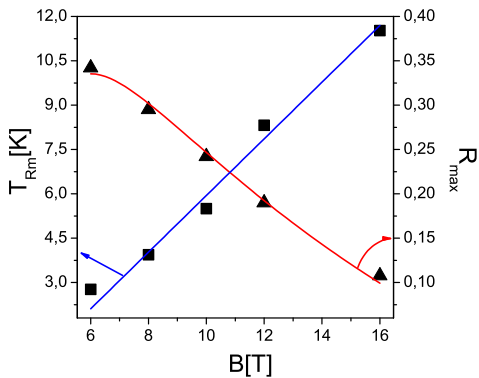}
\end {center}\vspace{-1.0cm}
\caption {The peak temperatures $T_{\rm Rm}$ (squares) and the peak
values $R_{\rm max}$ (triangles) versus magnetic field $B$ extracted
from measurements of MR \cite{pag}. The solid lines represent our
calculations based on eqs. \eqref{RTB1} and \eqref{RTB2}.}\label{TB}
\end{figure}

Before discussing  the magnetoresistance $\rho_{mr}(B,T)$ given by
eq. \eqref{HC23}, we consider the magnetic field dependencies of
both the peak value $R_{\rm max}(B)$ of MR and peak temperature
$T_{\rm Rm}(B)$ at which $R_{\rm max}(B)$ takes place. It is
possible to use eq. \eqref{HC26} which relates the position and
value of the peak with the function $M^*(B,T)$. To do this, we have
to take into account the classical contribution
$\Delta\rho_{L}(B,T)$ to MR and the residual resistance $\rho_0$
which prevent $T_{\rm Rm}(B)$ from vanishing and makes $R_{\rm
max}(B)$ finite at $B\to B_{c0}$. Therefore, MR is a continuous
function at the quantum critical point $B_{c0}$ in contrast to
$M^*(B,T)$ which peak value diverges and the peak temperature tends
to zero at the point as it follows from eqs. \eqref{B32} and
\eqref{YTB}. As a result, we have to substitute $B_{c}$ for $B_{c0}$
and take $B_{c}$ as a parameter. Upon modifying eq. \eqref{HC26} by
taking into account $\Delta\rho_{L}(B,T)$ and $\rho_0$, we obtain
\begin{equation}\label{RTB1}
T_{\rm Rm}(B)\simeq b_1(B-B_{c}),\end{equation}
\begin{equation}\label{RTB2} R_{\rm max}(B)\simeq
\frac{b_2(B-B_{c})^{-1/3}-1}{b_3(B-B_{c})^{-1}+1}.
\end{equation} Here $b_1$, $b_2$, $b_3$ and $B_c$ are fitting
parameters. It is pertinent to note that when deriving eq.
\eqref{RTB2}, eq. \eqref{RTB1} was employed in substituting
$(B-B_c)$ for $T$. Then, eqs. \eqref{RTB1} and \eqref{RTB2} are not
valid at $B<B_{c0}$ when the HF metal obtains both the
antiferromagnetic order and LFL behavior. In fig. \ref{TB}, we show
the field dependence of both $T_{\rm Rm}$ and $R_{\rm max}$,
extracted from measurements of MR \cite{pag}. It is seen that both
$T_{\rm Rm}$ and $R_{\rm max}$ are well described by eqs.
\eqref{RTB1} and \eqref{RTB2} with $B_{c}=$3.8 T. We note that this
value of $B_{c}$ is in good agreement with observations obtained
from the $B-T$ phase diagram of \chem{CeCoIn_5}, see fig. 3 of Ref.
\cite{pag}.

\begin{figure} [! ht]
\begin{center}
\includegraphics [width=0.47\textwidth]{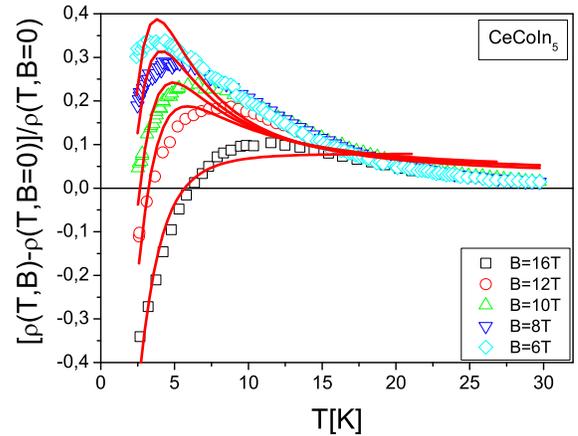}
\end {center}\vspace{-1.0cm}
\caption {MR versus temperature $T$ as a function of magnetic field
$B$. The experimental data on MR were collected on \chem{CeCoIn_5}
at fixed magnetic field $B$ \cite{pag} shown in the right bottom
corner of the figure. The solid lines represent our calculations,
eq. \eqref{UN2} is used to fit the effective mass entering eq.
\eqref{HC27}.}\label{MRT}
\end{figure}

To calculate MR $\rho_{mr}(B,T)$, we apply eqs. \eqref{HC27} to
describe the universal behavior of MR, eq. \eqref{UN2} to describe
the behavior of the effective mass and eqs. \eqref{RTB1} and
\eqref{RTB2} to assign the absolute values to MR. Figure \ref{MRT}
shows the calculated MR versus temperature as a function of magnetic
field $B$, together with the facts taken from Ref. \cite{pag}. We
recall that the contributions coming from $\Delta\rho_{L}(B,T)$ and
$\rho_0$ were omitted. As seen from fig. \ref{MRT}, our description
of the facts is quite good and we conclude that main contribution to
MR comes from the dependence of the effective mass on the applied
magnetic magnetic field $B$.

In summary, we have performed a study of MR of the HF metal
\chem{CeCoIn_5} within the framework of the fermion condensation
quantum phase transition. Obtained results are in good agreement
with facts and have allowed us to reveal new scaling behavior of MR.


\begin{thebibliography}{199}

\bibitem{pag}\Name{Paglione J. \etal} \REVIEW{Phys. Rev.
Lett.}{91}{2003}{246405}.

\bibitem{mal}
\Name{Malinowski A.} \REVIEW{Phys. Rev. B}{72}{2005}{184506}.

\bibitem{zim}\Name{Ziman J. M.}
\Book{Electrons and Phonons} \Publ{Oxford University Press, Oxford}
\Year{1960}.

\bibitem{daybell}\Name{Daybell M. D. \and Steyert W. A.} \REVIEW{Phys. Rev.
Lett.}{18}{1967}{398}.

\bibitem{voj} \Name{Vojta M.} \REVIEW{Rep. Prog. Phys.}{66}{2003}{2069}.

\bibitem{loh}\Name{L\"ohneysen H.v. \etal}\REVIEW{Rev. Mod. Phys.}
{79}{2007}{1015}.

\bibitem{si} \Name{Gegenwart P., Si Q. \and Steglich F.} \REVIEW{Nature Phys.} {4}{2008}{186}.

\bibitem{kadw} \Name{Kadowaki K. \and Woods S.B.}   \REVIEW{
Solid State Commun.} {58}{1986}{507}.

\bibitem{tky}   \Name{Tsujii N., Kontani H. \and  Yoshimura K.}
\REVIEW {Phys. Rev. Lett.} {94}{2005}{057201}.

\bibitem{pag1}\Name{Paglione J. \etal} \REVIEW{Phys. Rev.
Lett.}{97}{2006}{106606}.

\bibitem{steg} \Name{Gegenwart P. \etal} \REVIEW{Science}{315}{2007}{969}.

\bibitem{ckz} \Name{Khodel V. A., Zverev M. V. \and Yakovenko V. M.}
\REVIEW{Phys. Rev. Lett.}{95}{2005}{236402}.

\bibitem{khod} \Name{Clark J. W., Khodel V. A. \and Zverev M. V.}
\REVIEW{Phys. Rev. B}{71}{2005}{012401}.

\bibitem{obz} \Name{Shaginyan V. R., Amusia M. Ya. \and Popov K. G.} \REVIEW{
Physics-Upsekhi}{50}{2007}{563}.

\bibitem{shag1}\Name{Shaginyan V. R. \etal} \REVIEW{Europhys.
Lett.}{76}{2006}{898}.

\bibitem{shag2} \Name{Shaginyan V. R., Popov K. G. \and Stephanovich V. A.} \REVIEW{Europhys.
Lett.}{79}{2007}{47001}.

\bibitem{shag3} \Name{Shaginyan V. R. \etal} \REVIEW{Phys. Rev.
Lett.}{100}{2008}{096406}.

\bibitem{khs} \Name{Khodel V.A. \and Shaginyan V. R.} \REVIEW{JETP Lett}
{51}{1990}{553} .

\bibitem{ams} \Name{Amusia M. Ya. \and Shaginyan V. R.} \REVIEW{Phys. Rev.
B}{63}{2001}{224507}.

\bibitem{volovik} \Name{Volovik G.E.} \Book{Quantum Phase Transitions from
Topology in Momentum Space} \Publ{Lect. Notes Phys.} 718, pp. 31-73,
\Year{2007}.

\bibitem{land}\Name{Lifshitz E. M. \and Pitaevskii L. P.}
\Book{Statistical Physics, Part 2} \Publ{Butterworth-Heinemann,
Oxford} \Year{1999}.

\bibitem{takah}
\Name{Takahashi  D. \etal} \REVIEW{Phys. Rev.
B}{67}{2003}{180407(R)}.

\bibitem{he3} \Name{Neumann M., Ny\'{e}ki J. \and Saunders J.} \REVIEW{Science}
{317}{2007}1356.

\bibitem{pikul} \Name{Pikul A.P. \etal} \REVIEW{J. Phys. Condens. Matter} {18}{2006}{L535}.

\bibitem{shag_mr} \Name{Shaginyan V. R.} \REVIEW{JETP Lett.}{77}{2003}{178}.

\end{thebibliography}
\end{document}